\documentclass[runningheads]{llncs}
\usepackage[T1]{fontenc}
\usepackage{xcolor}
\usepackage{graphicx}
\usepackage{hyperref}
\usepackage{pifont}
\usepackage{multirow}

\newcommand{\cmark}{\ding{51}} % Check mark
\newcommand{\xmark}{\ding{55}} % Cross mark
\begin{document}
\title{QuickDraw: Fast Visualization, Analysis and Active Learning for Medical Image Segmentation}
\titlerunning{QuickDraw}
\author{Daniel Syomichev$^{1,}$\thanks{Equal Contribution, $^\dagger$ Equal Advising}, Padmini Gopinath$^{1,\star}$, Guang-Lin Wei$^1$, Eric Chang$^1$, Ian Gordon$^1$, Amanuel Seifu$^1$, Rahul Pemmaraju$^{2,\dagger}$, Neehar Peri$^{3,\dagger}$, James Purtilo$^{1,\dagger}$
}
\authorrunning{D Syomichev and P Gopinath et al.}
\institute{University of Maryland$^1$, Rutgers University$^2$, Carnegie Mellon University$^3$}
\maketitle              % typeset the header of the contribution
\begin{abstract}
Analyzing CT scans, MRIs and X-rays is pivotal in diagnosing and treating diseases. However, detecting and identifying abnormalities from such medical images is a time-intensive process that requires expert analysis and is prone to interobserver variability. To mitigate such issues, machine learning-based models have been introduced to automate and significantly reduce the cost of image segmentation.  Despite significant advances in medical image analysis in recent years, many of the latest models are never applied in clinical settings because state-of-the-art models do not easily interface with existing medical image viewers. To address these limitations, we propose QuickDraw, an open-source framework for medical image visualization and analysis that allows users to upload DICOM images and run off-the-shelf models to generate 3D segmentation masks. In addition, our tool allows users to edit, export, and evaluate segmentation masks to iteratively improve state-of-the-art models through active learning. In this paper, we detail the design of our tool and present survey results that highlight the usability of our software. Notably, we find that QuickDraw reduces the time to manually segment a CT scan from four hours to six minutes and reduces machine learning-assisted segmentation time by 10\% compared to prior work. Our code and documentation are available on \href{https://github.com/qd-seg/quickdraw}{GitHub}. 

\keywords{Medical Image Analysis \and Active Learning \and Segmentation}
\end{abstract}
\section{Introduction}
\begin{figure}[t]
    \centering
    \includegraphics[width=\linewidth]{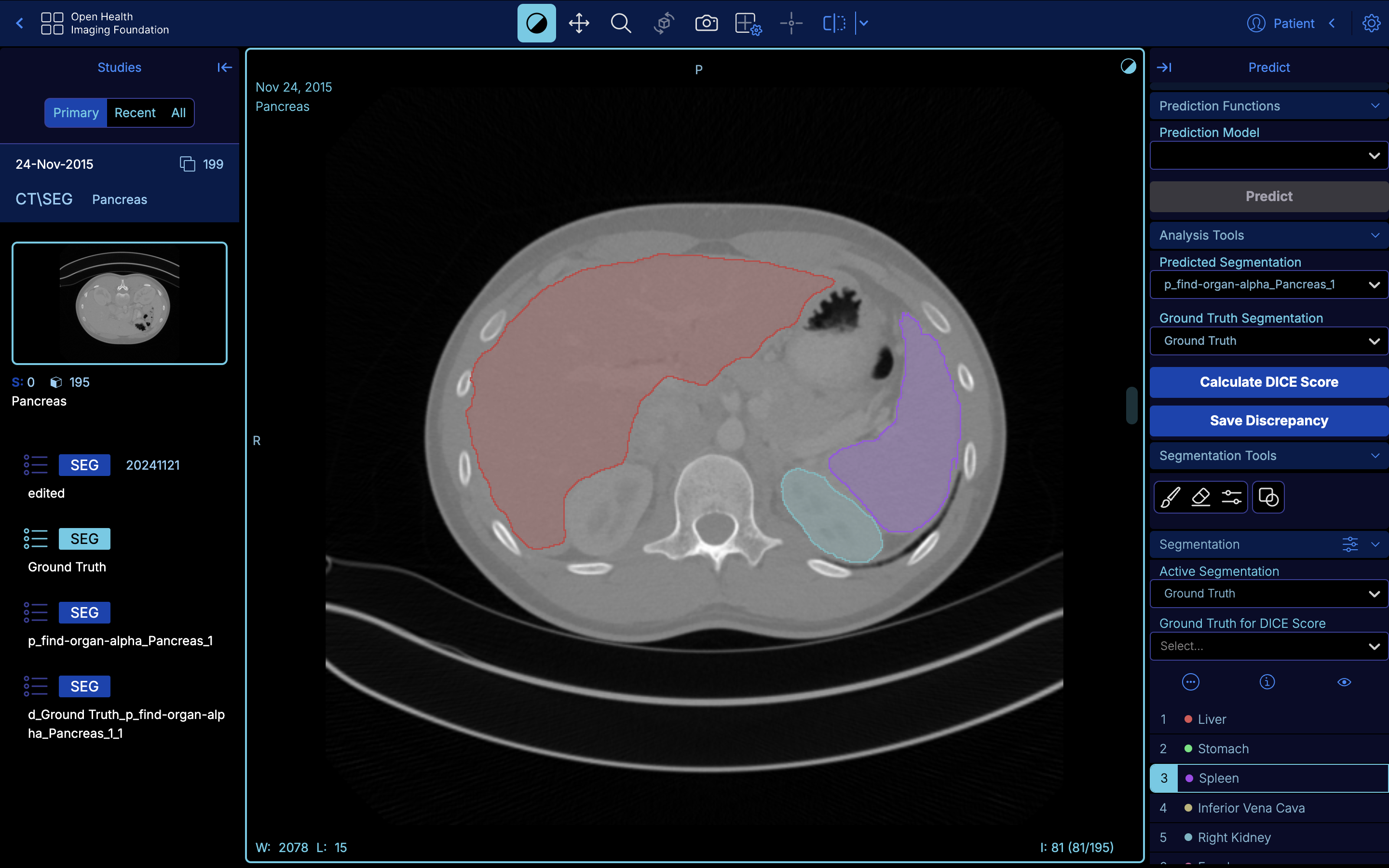}
    \caption{We present QuickDraw, an open-source medical image segmentation visualization, analysis, and annotation tool that allows users to add, delete, and modify manually drawn segmentation masks or machine learning-based predictions to support active learning.}
    \label{fig:teaser}
\end{figure}

Medical image analysis of CT scans, X-rays, and MRIs plays a crucial role in diagnosis and treatment. However, manually analyzing such medical images can be time-consuming and often requires domain-specific expertise. Therefore, machine learning-based models have been introduced to improve diagnostic speed and accuracy. 

{\bf Status Quo.} Recent advances in learning-based segmentation demonstrate high throughput and can often match the accuracy of experienced medical experts \cite{Cardoso2022,Riem2023}. However, training such data-driven models requires large-scale annotated datasets and curating high-quality training data can take up to four hours per 3D scan \cite{vanSluis2023}. Moreover, many state-of-the-art models are never integrated into clinical or research workflows because existing models do not easily interface with existing medical image viewers. To the best of our knowledge, no existing software solution provides a unified interface for model inference, evaluation, and active learning. Instead, existing workflows are often fragmented between different platforms for visualization (such as 3D Slicer \cite{Fedorov2012} and OHIF Viewer \cite{Ziegler2020}), model predictions, and image analysis. Existing methods are time-consuming, error-prone, and are often inaccessible to users with limited technical expertise or computing resources.

{\bf Contributions.} In this paper, we introduce QuickDraw, an open-source framework for medical image visualization and machine learning-assisted segmentation. QuickDraw enables experts to upload medical images, run inference with pre-trained models, and generate multi-organ segmentation predictions within a single workflow (cf. Fig \ref{fig:teaser}). Our tool includes built-in evaluation metrics, such as the DICE score, and highlights discrepancies between model predictions and ground-truth segmentation masks. Recognizing that model predictions may not always be perfect, QuickDraw allows users to refine predicted masks and export these updated masks to downstream active learning pipelines, improving model accuracy over time. We evaluate the effectiveness of our tool through several small-scale user studies and find that QuickDraw is intuitive to use and highly effective in accelerating medical image segmentation.

\section{Related Works}
In this section, we present a brief overview of existing image viewers, machine learning frameworks, and active learning methods for medical image analysis.

\textbf{Medical Image Viewers and ML Frameworks.}  Although existing tools support a subset of common use cases such as medical image visualization, model-based segmentation, model evaluation, and 3D image annotation, to the best of our knowledge, no existing tool supports end-to-end medical image analysis. Software packages such as 3D Slicer \cite{Fedorov2012}, OHIF Viewer \cite{Ziegler2020}, and ITK-SNAP \cite{Yushkevich2017} facilitate the visualization and manual annotation of medical images and provide an interface for users to interact with DICOM files. Machine learning frameworks such as MONAI \cite{Cardoso2022}, MedSAM \cite{Ma2024}, MedSegDiff-V2 \cite{Wu2023}, U-Net \cite{Ronneberger2015}, nnU-Net \cite{Isensee2018}, DUCK-Net \cite{Dumitru2023}, and V-Net \cite{Milletari2016} simplify model training \cite{Cardoso2022,pemmaraju2024semi,Shi2024}. However, previous work \cite{Perez-Garcia2024,Jungo2024,Ronneberger2015,Hussain2024} relies on external tools to flag low-quality predictions and refine segmentation masks for active learning.  Unlike existing tools, QuickDraw combines visualization, learning-based segmentation, model evaluation, and prediction re-annotation in a single workflow. Our user studies suggest that QuickDraw streamlines medical image analysis, encourages iterative model improvement, and improves efficiency. 

\textbf{Active Learning Methods for Medical Image Analysis.} Although state-of-the-art models achieve high accuracy on standard benchmarks, model predictions may not always be perfect on out-of-distribution data. Active learning has emerged as a promising method for iteratively improving model performance. By selectively re-annotating and training on a targeted subset of the most informative examples, active learning can significantly improve model accuracy while reducing the need for large-scale annotated data  \cite{chen2023rebound}. A key challenge in active learning is identifying the most informative examples for annotation. Methods such as Predictive Accuracy-Based Active Learning (PAAL) \cite{Shi20241}, Adaptive Superpixel-Based Active Learning \cite{Kim2023}, Stochastic Batch-Based Active Learning \cite{Gaillochet2023}, and Selective Uncertainty-Based Active Learning \cite{Ma20241} have introduced  uncertainty-based  \cite{Nath2021,Ma20241,Xie2022,Sourati2019}, diversity-based \cite{Gaillochet2023}, and hybrid approaches that combine both uncertainty and diversity methods \cite{Shi20241,Smailagic2018,Liu2023} to identify informative samples. Despite these advancements, active learning methods often rely on external tools to refine predictions or process annotated data. QuickDraw supports active learning by allowing users to edit low-quality predictions, export corrected outputs, and re-evaluate model performance to facilitate iterative model improvement.

\section{Method}
Medical image segmentation plays a crucial role in identifying and highlighting regions that may be relevant to a diagnosis. These regions of interest (ROIs) can include organs, tumors, or other diagnostically relevant structures. QuickDraw is designed to simplify the visualization and analysis of deep-learning model predictions.

\begin{figure}[t]
    \centering
    \includegraphics[width=\textwidth]{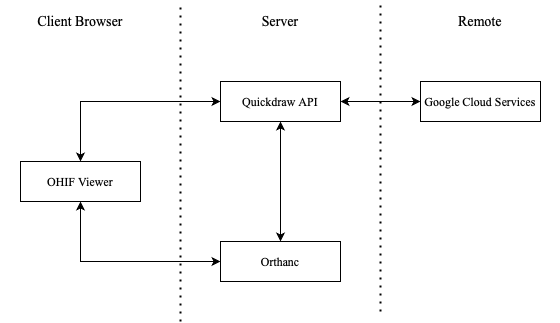}
    \caption{QuickDraw extends the OHIF Viewer by connecting an Orthanc server to store DICOM images and Google Cloud Services (GCS) to run model inference.}
    \label{fig:prediction}
\end{figure}

\textbf{Model Management.}
We use Google Cloud Services (GCS) for its flexible and efficient approach to storing and running deep learning models. Using command-line interface (CLI) scripts, we can upload a Docker container to GCS and deploy it on demand for model inference. When an inference request is triggered, a virtual machine (VM) is spun up and run for the duration of the task. Importantly, this VM instance is automatically suspended after each inference call, minimizing hardware overhead and significantly reducing costs.

\textbf{Visualization and Inference.}
The Open Health Imaging Foundation (OHIF) Viewer is a web-based medical image viewer that supports various image types (e.g. DICOM) and their corresponding segmentation masks (e.g. RTStructs). It displays images in a layer-based configuration and includes a basic set of segmentation editing tools. Users can create segmentation masks from scratch with any number of regions of interest (ROIs) or refine existing masks for greater accuracy. Each ROI is uniquely colored with adjustable opacity. QuickDraw enhances OHIF with functionality to run inference on models uploaded to GCS. Once an image is loaded into the viewer, users can select a model from a drop-down menu and initiate an inference call.  After model inference is complete, users can refine the output to improve segmentation precision. This workflow significantly reduces the time required to annotate an image compared to manual segmentation from scratch.

\textbf{Image Analysis.}
QuickDraw provides tools for both quantitative and qualitative analysis of predicted and ground truth segmentation masks. The OHIF Viewer can load and overlay multiple segmentation masks, making imprecise segmentation masks apparent through misaligned borders. To further highlight these differences, users can generate a discrepancy map (e.g. with an XOR bit mask that isolates non-matching regions between predicted and ground truth masks). In addition, a DICE coefficient can be calculated for each ROI. These analysis tools help users identify prediction errors and assess the reliability of model outputs.

\textbf{Software Architecture.}
As shown in Figure \ref{fig:prediction}, the OHIF Viewer serves as the primary user interface, offering a web-based viewer that allows multiple concurrent users to visualize DICOM images. To store and organize DICOM records, QuickDraw utilizes the Orthanc DICOM server. Through its REST API, a Python-based intermediary server retrieves images and uploads them to a GCS VM for prediction. Once the prediction is complete, the resulting segmentation is sent back to the Orthanc server. All medical images are deleted from the server after inference to preserve patient data privacy. QuickDraw further uses the GCS Artifact Registry to store containerized models and the GCS Compute Engine to allocate resources for model inference.

\begin{table}[t]
    \centering
    \begin{tabular}{|p{9cm}|c|}
        \hline
        \textbf{Question} & \textbf{Average Response} \\
        \hline
        It was easy to navigate the user interface & 3.25 \\
        \hline
        It was easy to use the segmentation tool & 3.25 \\
        \hline
        You're satisfied with the overall performance and responsiveness of this application & 4.5 \\
        \hline
        The tool reduced the time or effort required to complete your tasks compared to existing workflow & 3.75 \\
        \hline
        The DICE Score calculation is useful for evaluating the accuracy of a model's prediction & 4.5 \\
        \hline
        You're likely to adopt this tool in your daily workflow & 3.75 \\
        \hline
        You would recommend this product to a peer & 4.25 \\
        \hline
    \end{tabular}
    \caption{{\bf Survey Results on Tool Usability.} We ask four participants with prior experience in medical imaging to rate the usability of QuickDraw on a likert scale: Strongly Disagree (1), Disagree (2), Neutral (3), Agree (4), Strongly Agree (5). A high recommendation score indicates strong user approval.}
    \label{tab:survey_results}
\end{table}

\section{Experiments}
In this section, we present the results of our experiments evaluating the effectiveness of QuickDraw. 

\textbf{Ease of Use.} 
We evaluate QuickDraw's usability through timed tasks and user surveys. Eight participants -- four with and four without medical imaging experience -- completed tasks under guided supervision. The four participants with medical imaging experience also rated their experience on a five-point likert scale. Survey results (cf. Table \ref{tab:survey_results}) indicate that users find the interface moderately easy to navigate and the segmentation tool intuitive to use. Participants scored QuickDraw's performance and responsiveness highly, and noted that the DICE score calculation was particularly useful. While our workflow efficiency experiment (cf. Table \ref{tab:segmentation_results}) suggests advantages over existing methods, further refinement may be needed for full adoption.

Table \ref{tab:segmentation_results} shows that non-experienced users significantly reduced task completion time on their second attempt, indicating QuickDraw’s shallow learning curve. Participants found that the main challenge was in modifying segmentation masks. In particular, participants found it difficult to adjust the annotation brush size and use the sphere feature. Despite initial difficulty, user efficiency improved considerably in the second trial. Experienced users had faster segmentation times and lower workflow completion times while maintaining high DICE scores, highlighting the efficiency gained from prior medical imaging experience. Model-based predictions further streamlined annotation workflows, further reducing annotation time.

\textbf{Segmentation Time and Dice Score Evaluation.} To assess segmentation efficiency, users modified model predictions to match a provided ground-truth mask. Results show that model-assisted segmentation was significantly faster than manual segmentation, with improved DICE scores after user refinements. QuickDraw’s DICE score also helps assess segmentation quality by correlating accuracy with completion time. For example, a high DICE score in a short time (Participant 4) suggests increased efficiency, while a longer time with a high score (Participant 1) indicates that the participant is more meticulous. These correlations provide additional insights into user interactions.

{
\setlength{\tabcolsep}{2mm}
\begin{table}[t]
\scalebox{0.55}{
\begin{tabular}{|c|c|cc|cc|cc|c|cc|c|}
\hline
\multirow{2}{*}{{\bf Participant}} & \multirow{2}{*}{{\bf 
 Experience}} & \multicolumn{2}{c|}{{\bf Final Dice Score}}  & \multicolumn{2}{c|}{{\bf $\Delta$ Dice Score}}  & \multicolumn{2}{c|}{{\bf Seg. Time (s)}}     &\multirow{2}{*}{{\bf $\Delta$ Seg. Time (s)}} & \multicolumn{2}{c|}{{\bf Total Time (m)}}   &\multirow{2}{*}{{\bf $\Delta$ Total Time (m)}} \\ \cline{3-8} \cline{10-11} 
                                &                              & \multicolumn{1}{c|}{Trial 1} & Trial 2 & \multicolumn{1}{c|}{Trial 1} & Trial 2 & \multicolumn{1}{c|}{Trial 1} & Trial 2 &   & \multicolumn{1}{c|}{Trial 1} & Trial 2 & \\ \hline
                                
    1  &   \cmark    & \multicolumn{1}{c|}{0.95}        &    -     & \multicolumn{1}{c|}{0.17}        &     -    & \multicolumn{1}{c|}{510}        &   -      &    -      & \multicolumn{1}{c|}{16:16}        &   -     & -  \\ 

    2 &   \cmark   & \multicolumn{1}{c|}{0.93}        &    -     & \multicolumn{1}{c|}{0.15}        &     -    & \multicolumn{1}{c|}{153}        &    -     &     -      & \multicolumn{1}{c|}{8:18}        &    -     & -\\

    3  &   \cmark   & \multicolumn{1}{c|}{0.84}        &    -     & \multicolumn{1}{c|}{0.06}        &    -     & \multicolumn{1}{c|}{135}        &     -    &     -      & \multicolumn{1}{c|}{3:55}        &   -      & -\\ 

    4    &   \cmark   & \multicolumn{1}{c|}{0.95}        &    -     & \multicolumn{1}{c|}{0.17}        &    -     & \multicolumn{1}{c|}{88}        &     -    &     -      & \multicolumn{1}{c|}{4:53}        &    -     & - \\

    5   &   \xmark   & \multicolumn{1}{c|}{0.93}        &   0.93      & \multicolumn{1}{c|}{0.15}        &   0.15      & \multicolumn{1}{c|}{302}        &    295     &   7        & \multicolumn{1}{c|}{10:23}        &    6:22 &   4:01  \\ 

    6    &   \xmark   & \multicolumn{1}{c|}{0.89}        &   0.95     & \multicolumn{1}{c|}{0.11}        &    0.17     & \multicolumn{1}{c|}{448}        &    218     &    230       & \multicolumn{1}{c|}{15:12}        &   8:07 &  7:05   \\

    7    &   \xmark   & \multicolumn{1}{c|}{0.95}        &   0.94     & \multicolumn{1}{c|}{0.17}        &    0.16     & \multicolumn{1}{c|}{350}        &   340      &    10       & \multicolumn{1}{c|}{10:33}        &    8:43  &  1:50 \\ 

    8     &  \xmark  & \multicolumn{1}{c|}{0.96}        &   0.96     & \multicolumn{1}{c|}{0.18}        &    0.18     & \multicolumn{1}{c|}{884}        &   464      &     420      & \multicolumn{1}{c|}{17:11}        &   11:56   &  5:15 \\ 
                                \hline

\end{tabular}}
\caption{\textbf{Segmentation Performance and Time Metrics.} We evaluate eight participants (four with prior medical imaging experience, and four without prior experience) on their ability to quickly adapt to QuickDraw's interface and modify model predictions. Notably, we find that all participants improve initial model predictions (Initial DICE Score = 0.78). Further, inexperienced participants are able to significantly reduce their time to completion in their second attempt, suggesting that QuickDraw has a shallow learning curve.}
    \label{tab:segmentation_results}
\end{table}
}

\section{Conclusion}
This paper introduces QuickDraw, an open-source framework for medical image visualization and machine learning-assisted segmentation. QuickDraw streamlines standard image analysis workflows by allowing experts to upload medical images, run inference with pre-trained models, and analyze multi-organ segmentation predictions. Since model predictions are not always perfect, QuickDraw also allows users to refine segmentation masks, export updated predictions, and integrate them into active learning pipelines to improve model accuracy over time. Through multiple user studies, we find that QuickDraw is both intuitive and highly efficient in accelerating medical image segmentation. More broadly, we argue that by lowering barriers to integrating and testing diverse models, QuickDraw can improve clinical workflows, enable researchers to experiment with state-of-the-art models, and expedite the validation of new tools for clinical use, ultimately improving patient outcomes.

\section{Acknowledgments}
This work was supported in part by funding from the NSF GRFP (Grant No. DGE2140739).

\newpage 

\bibliographystyle{splncs04}
\bibliography{main}

\end{document}